\documentclass[letter]{aa}
\usepackage{txfonts}
\usepackage{natbib}
\usepackage{graphicx}
\bibpunct{(}{)}{;}{a}{}{,} 

\newcommand{\ind}[1]{_{\mathrm{#1}}}
\newcommand{\diff}{\mathrm{d}}
\def\numax{\nu\ind{max}}

\def\Dnu{\Delta\nu}
\def\dnumoy{\langle\Delta\nu\rangle}
\def\dnumoy{\Dnu}
\def\dnug{\Dnu\ind{guess}}
\def\dd{\delta\nu}

\begin{document}
\title{The universal red-giant oscillation pattern}
\subtitle{an automated determination with CoRoT data}
\titlerunning{The universal red-giant oscillation pattern}
\author{B. Mosser\inst{1}\and
K. Belkacem\inst{2,1} \and M.J. Goupil\inst{1} \and
E.Michel\inst{1}\and Y. Elsworth\inst{3}\and C. Barban\inst{1}\and
T. Kallinger\inst{4,5}\and S. Hekker\inst{3,6}\and J. De
Ridder\inst{6}\and R. Samadi\inst{1}\and F. Baudin\inst{7}\and
F.J.G. Pinheiro\inst{1,8}\and M. Auvergne\inst{1}\and A.
Baglin\inst{1}\and C. Catala\inst{1}}

\offprints{B. Mosser}

\institute{LESIA, CNRS, Universit\'e Pierre et Marie Curie, Universit\'e Denis Diderot,
Observatoire de Paris, 92195 Meudon cedex, France; \email{benoit.mosser@obspm.fr}
\and
Institut d'Astrophysique et de
G\'eophysique, Universit\'e de Li\`ege, All\'ee du 6 Ao\^ut, 17
B-4000 Li\`ege, Belgium
\and
School of Physics and Astronomy, University of Birmingham,
Edgbaston, Birmingham B15 2TT, United Kingdom
\and
Institute for Astronomy (IfA), University of Vienna,
T\"urkenschanzstrasse 17, 1180 Vienna, Austria
\and
Department of Physics and Astronomy, University of British Columbia,
Vancouver, BC V6T 1Z1, Canada
\and
Instituut voor Sterrenkunde, K. U. Leuven, Celestijnenlaan 200D, 3001 Leuven, Belgium
\and
Institut d'Astrophysique Spatiale, UMR 8617, Universit\'e Paris XI,
B\^atiment 121, 91405 Orsay Cedex, France
\and
Centro de F\'\i sica Computacional, Department of Physics,
University of Coimbra, 3004-516 Coimbra, Portugal
}

\abstract{}
{The CoRoT and Kepler satellites have provided thousands of
red-giant oscillation spectra. The analysis of these spectra
requires efficient methods for identifying all eigenmode
parameters.}
{The assumption of new scaling laws allows us to construct a
theoretical oscillation pattern. We then obtain a highly precise
determination of the large separation by correlating the observed
patterns with this reference.
}%
{We demonstrate that this pattern is universal and are able to
unambiguously assign the eigenmode radial orders and angular
degrees. This solves one of the current outstanding problems of
asteroseismology hence allowing precise theoretical investigation
of red-giant interiors.} {}

\keywords{Stars: oscillations - Stars: interiors - Methods: data
analysis - Methods: analytical}

\maketitle

\voffset = 1.5cm
\section{Introduction\label{introduction}}

Red giants are evolved stars that have depleted the hydrogen in
their cores and are no longer able to generate energy from
core-hydrogen burning. The physical processes taking place in
their interiors are currently rather poorly understood.
Observations with the space-borne mission CoRoT have revealed the
oscillation pattern \citep{2009Natur.459..398D} of many of these
stars, which is a crucial step on the route to probing their
internal structure. Before the advent of the CoRoT data, complex
oscillation patterns were explained by short-lived modes of
oscillation \citep{2004SoPh..220..207S,2007A&A...468.1033B}.
The new era of the space-borne missions CoRoT and \emph{Kepler}
has dramatically increased the amount and quality of the available
asteroseismic data of red giants
\citep{2009A&A...506..465H,2010ApJ...713L.176B,2010A&A...517A..22M,huber2010}.
The analysis of the oscillation eigenmodes now allows seismic
inferences to be drawn about the internal structure.

The identification of the angular degree and radial order of the
eigenmodes represents a first and crucial step in an asteroseismic
analysis. The values of the eigenfrequencies  can be related to
the order and degree of a mode with the commonly used asymptotic
equation:
\begin{equation}
\nu_{n,\ell} = \left[ n+{\ell\over 2} + \varepsilon \right] \, \Delta\nu - \dd_{0\ell}
\label{tassoul}
\end{equation}
where $\nu_{n,\ell}$ is the eigenfrequency of a mode with radial
order $n$ and angular degree $\ell$; $\Delta\nu$ is the mean value
of the large separation ($\dnumoy \simeq \nu_{n+1,\ell} -
\nu_{n,\ell}$), and $\dd_{0\ell}$ is a second-order term, or small
separation, dependent  on the mode degree. This form, which is
similar to the original expression developed for the Sun 30 years
ago \citep{1980ApJS...43..469T}, is useful for analysing the
observations and performing the complete mode identification. It
assumes $\dd_{00}$ equals 0. The parameter $\varepsilon$ comprises
two parts: the offset due to the mode propagation in the
upper-most layers of the star, and the second-order term of the
asymptotic approximation which is sensitive to the gradient of
sound speed in the stellar interior.

In the absence of accurate determinations of the individual mode
frequencies, the global seismic parameters used in  the asymptotic
expression above are important indicators of the physical
parameters of the star. The large separation gives a measure of
the mean stellar density; the small separation $\dd_{0\ell}$
describes the stratification of the central regions.
Unfortunately, the  methods currently used to determine the global
oscillation parameters suffer from various sources of uncertainty
\citep{2009A&A...506..465H,2009CoAst.160...74H,2010A&A...511A..46M}.
First, the stochastic excitation of the modes gives rise to
variability in the amplitudes, resulting in an apparently
irregular comb structure; second, the finite mode lifetime blurs
the estimates of the eigenfrequencies; third, estimates are
affected by the stellar noise and granulation signal superimposed
on the oscillations \citep{2009A&A...508..877M}. In fact,
simulations have shown that the impact of realization noise on the
measurement of the large separation $\dnumoy$,  can be much larger
than the background noise for red giants
\citep{2010arXiv1008.2959H}.

In an analysis of a sub-sample of Kepler red giants,
\cite{huber2010} have shown the regularity of the oscillation
spectra of such stars. In this paper, we show that all red giants
have a regular pattern, as modelled recently by
\cite{2010ApJ...721L.182M}. We propose a method which allows us to
tag all the modes with their appropriate radial order and angular
degree, regardless of the presence of the perturbing effects
described above.

\section{Method\label{method}}

The method to mitigate the effects of realization noise uses  Eq.
\ref{tassoul} in a dimensionless form:
\begin{equation}
{\nu_{n,\ell} \over \dnumoy} = n+{\ell\over 2}  + \varepsilon (\dnumoy) - d_{0\ell} (\dnumoy) .
\label{tassoul_m}
\end{equation}
In a departure from the previous practice, we have assumed that
$\varepsilon$ obeys a scaling law $\varepsilon = A + B \log
\dnumoy$, as derived from the observation of thousands of CoRoT
targets \citep{2010A&A...517A..22M} and as observed by
\cite{huber2010}. This is justified by the observation that
scaling laws apparently govern \emph{all} global asteroseismic
parameters
\citep{2009A&A...506..465H,2009MNRAS.400L..80S,2010ApJ...713L.176B,2010A&A...517A..22M}
and is equivalent to assume that the underlying physics of
$\varepsilon$ varies with the global stellar parameters.
As the mixed nature of dipole modes ($\ell=1$) is more pronounced,
we did not include them in the template, but only doublets
corresponding to the eigenmodes with even degrees ($\nu_{n-1,2}$
and $\nu_{n,0}$), with equal amplitudes. As a first guess, we set
the small separation $d_{02}$ at $- 0.14$ and then allowed it to
vary with the value of $\dnumoy$ according to the same
relationship as given for $\varepsilon$. For constructing the
peaks of the template, we have also  used the scaling laws of the
Gaussian excess power derived by \cite{2010A&A...517A..22M}. We
have assumed that the mode lifetime varies as $\dnumoy^{-1}$ and
have used mode widths equal to about a few percent of $\dnumoy$.
Finally, we stress that no background model is needed.

\begin{figure}
\includegraphics[width=8.2cm]{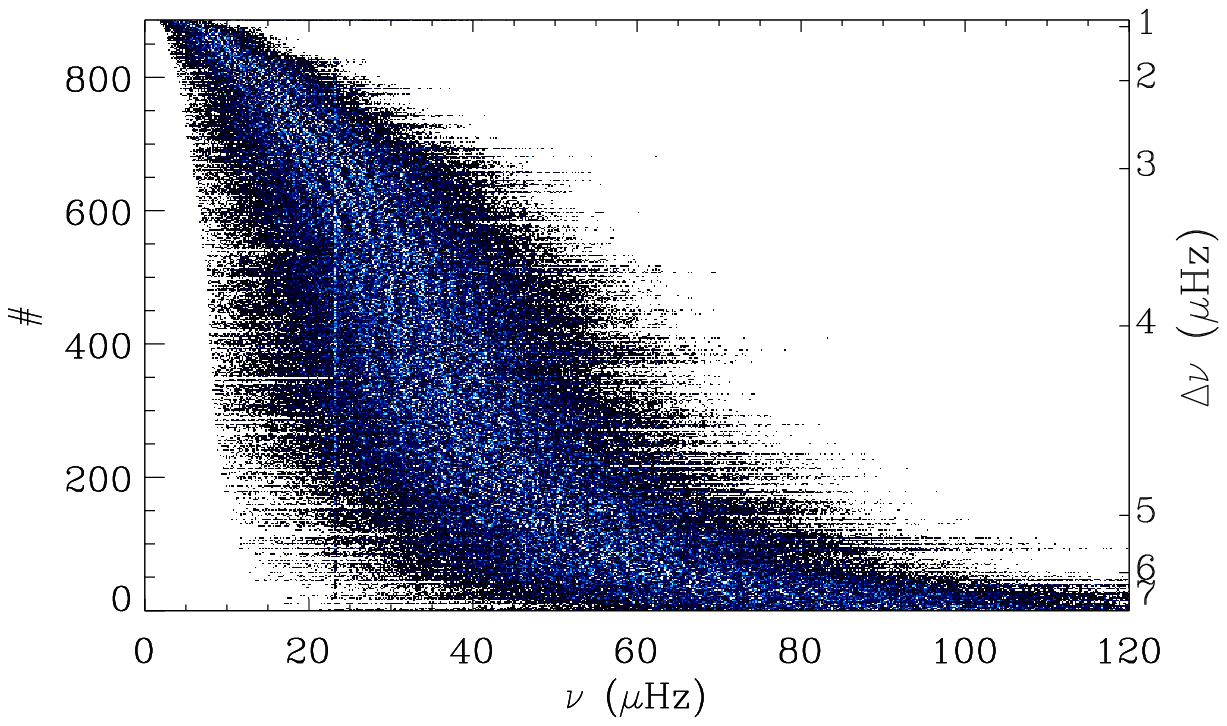}
\includegraphics[width=8.2cm]{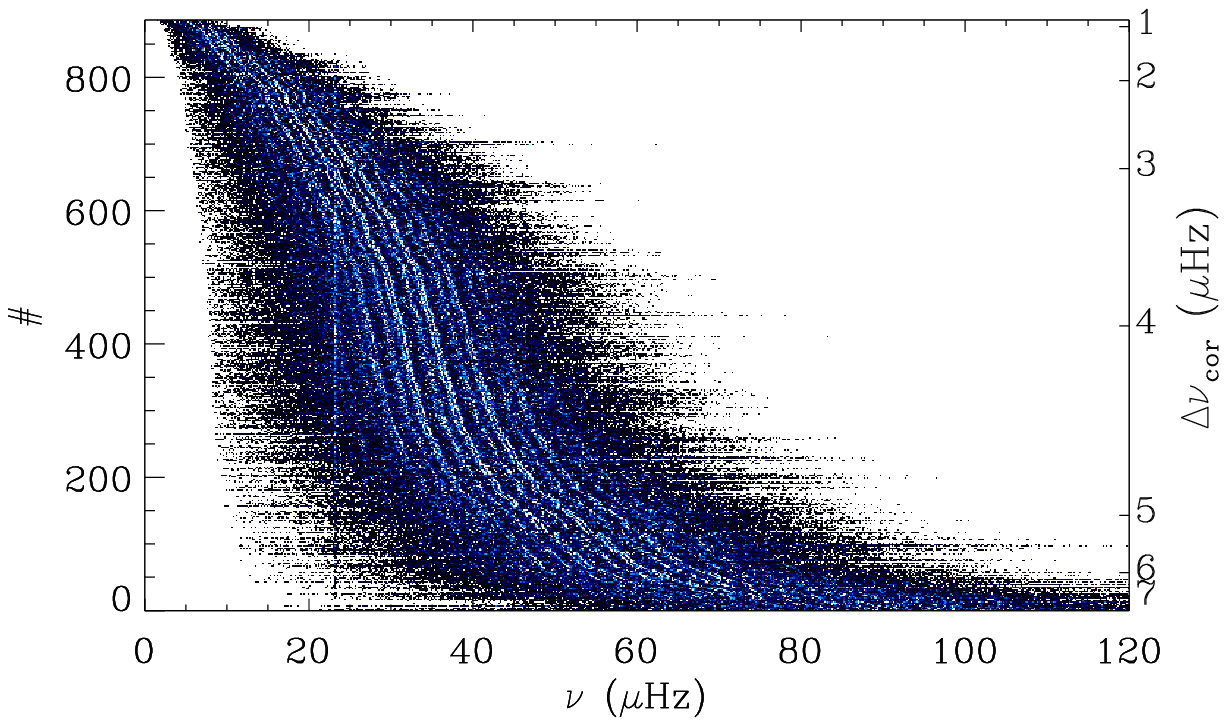}
\caption{CoRoT red giant power spectra stacked into an image after
sorting on the large separation. One line corresponds to one star.
{\bf a)} The first sorting is based on the large separation before
any correction. {\bf b)} The blurred aspect disappears once the
correction has been performed and reveals a clear comb-structure
common to all red giants. The vertical lines at 23.2\,$\mu$Hz are
the signature of the low-Earth orbit \citep{2009A&A...506..411A}.
\label{deter}}
\end{figure}

\begin{figure}
\includegraphics[width=8.7cm]{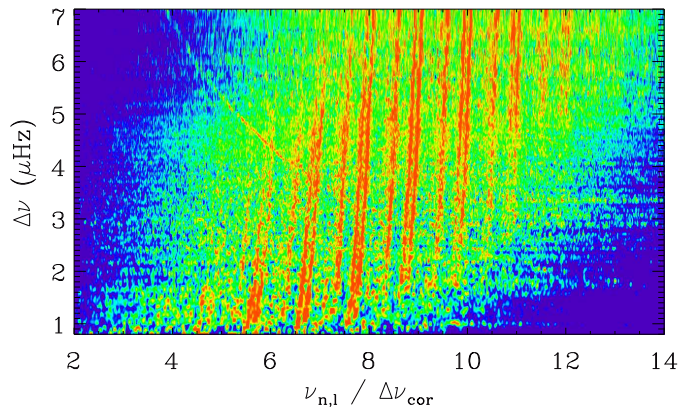}
\caption{The spectra presented in Fig. \ref{deter}b have been
rearranged in order to have the dimensionless frequency on the
abscissa and the mean large separation on the ordinate. They show
that the red-giant oscillation pattern is universal. The
hyperbolic branch is here the signature of the low-Earth orbit.
\label{universal}}
\end{figure}

The measurement of the large separation is performed in two steps.
First, an initial-guess value $\dnug$ of the large separation is
computed by an automated pipeline \citep{2009A&A...508..877M}.
This is used to form the initial synthetic template to correlate
with the real spectrum. The best correlation between the observed
and synthetic spectra provides then the corrected value of the
large separation. The template was iteratively adjusted by varying
its parameters to maximize the correlation.

In Fig.~\ref{deter}, we show the results obtained with all high
signal-to-noise CoRoT data \citep{2010A&A...517A..22M}. In both
cases the graphs show the spectra arranged in strips with the
colour representing the strength of the signal, as was done by
\citet{2010PASP..122..131G}. The spectra are sorted by increasing
large separation with the smallest large separation at the top of
the plots: in the upper plot we use the output from a conventional
pipeline and in the lower one we use the corrected
value. 
The remarkably regular structure within the oscillation spectra in
the lower plot reveals the signature of comb-like structure of the
asymptotic relationship in Eq.~(\ref{tassoul_m}) already reported
\citep{2009Natur.459..398D,2010A&A...509A..73C,2010ApJ...713L.176B,huber2010}.
Further, it validates the scaling law in $\varepsilon$ included in
the reference template. The global agreement of all high
signal-to-noise spectra of bright targets with the synthetic
pattern (Fig.~\ref{universal}) shows that these oscillation
patterns are homologous and that the red-giant oscillation pattern
is universal.

\begin{figure}
\includegraphics[width=8.8cm]{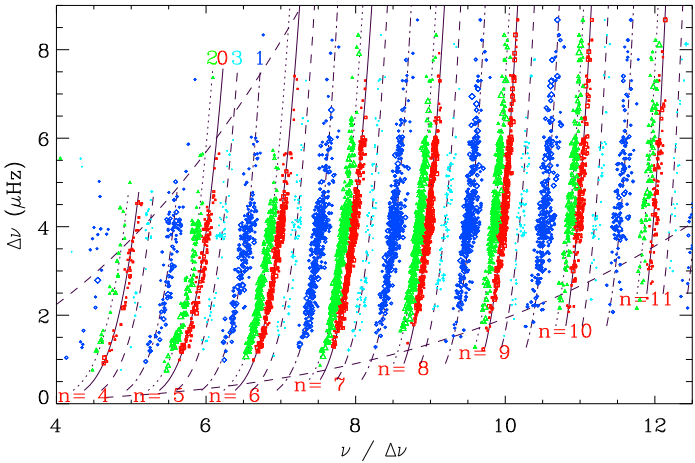}
\caption{Complete identification of the ridges, automatically
derived from the eigenfrequencies extracted with a
height-to-background ratio greater than 3. Each colour correspond
to a different mode degree (radial modes in red, dipole modes in
dark blue, $\ell=2$ modes in green, $\ell=3$ modes in light blue).
The solid grey lines superimposed on the ridges indicate for each
radial order the fits of $\varepsilon$ (with indication of the
radial order $n$). The fits of $d_{01}$, $d_{02}$ and $d_{03}$ are
superimposed on the respective ridges (respectively dash-dot, dot
and dash lines for $\ell = $1, 2 and 3).
The dark dashed lines, derived from the scaling law dealing with
the oscillation excess power, delineate the region where the modes
have noticeable amplitudes \citep{2010A&A...517A..22M}.
\label{identi}}
\end{figure}

We further found that the template is significantly improved if it
takes account of the linear dependence of the large separation in
frequency, expressed by the degree-dependent gradient $\alpha_\ell
= (\diff\log \dnumoy / \diff n)_\ell$:
\begin{equation}
{\nu_{n,\ell} \over \dnumoy} = n+{\ell\over 2}  + \varepsilon
(\dnumoy) - d_{0\ell} (\dnumoy) + {\alpha_\ell (\dnumoy) \over2}
\left( n - {\numax \over \dnumoy} \right)^2 \label{tassoul_mod}
\end{equation}
with $\numax$ the frequency of maximum oscillation amplitude. The
corrected values of $\dnumoy$ are derived from this template. The
values of the 12 free parameters that account for the variations
in frequency of the parameter $\varepsilon$, of the small
separations $d_{0\ell}$ and of the gradients $\alpha_\ell$ as
derived from the fits to more than 6\,000 eigenmodes
(Fig.~\ref{identi}) are given in Table \ref{fits}. The value of
$\varepsilon$, defined modulo 1, is fixed thanks to the
extrapolation to the Solar case. As noted by
\cite{2010A&A...509A..73C} and \cite{huber2010}, the small
separation $d_{01}$ is negative. All these fits are consistent
with Kepler results.


The average value of the correction from $\dnug$ to $\dnumoy$ is
of the order of 2.5\,\%. For the largest giants, with the smallest
values of $\dnumoy$ and the smallest ratio $\numax / \dnumoy$, the
correction can be as high as 6\,\%. The absolute difference
$|\dnug-\dnumoy|$ can be 10 times the estimate of the stellar
noise contribution. At low frequency, with an observing run not
much longer than the mode lifetimes \citep{baudin2010}, the
realization noise dominates the background noise and the mean
accuracy of the determination of $\dnumoy$ is uniform, at about
0.015\,$\mu$Hz. We are aware that any bias in the $\varepsilon
(\dnumoy)$ input relation will induce a bias in the results. We
therefore took care to insure the iterative process to be
unbiased.  Analysis of synthetic data indicates that the precision
gained with this method is of order 10 times better than that
obtained with conventional methods \citep{2010arXiv1008.2959H}.

\begin{table}
 \centering
 \caption{Fits of the ridges, for $\dnumoy$ expressed in $\mu$Hz}\label{fits}
 \begin{tabular}{llccc}
   \hline
\multicolumn{2}{c}{$\ell$} &\multicolumn{2}{c}{fit $A_\ell +B_\ell \log\dnumoy$}& gradient of $\Delta\nu_\ell$\\
                  &  & $A_\ell$           &  $B_\ell$         & $\alpha_\ell$   \\
\hline
0     &$\varepsilon$ & $ 0.634 \pm 0.008$ & $ 0.546 \pm 0.008$& $0.008\pm0.001$ \\
1     &$d_{01}$      & $-0.056 \pm 0.012$ & $-0.002 \pm 0.010$& $0.003\pm0.002$ \\
2     &$d_{02}$      & $ 0.131 \pm 0.008$ & $-0.033 \pm 0.009$& $0.005\pm0.001$ \\
3     &$d_{03}$      & $ 0.280 \pm 0.012$ & $0$               & $0.005\pm0.002$ \\
\hline
\end{tabular}
\end{table}

\section{Discussion\label{results}}

This new method based on a simple hypothesis and an automated
procedure removes any ambiguity on the identification of the modes
(Fig.~\ref{identi}), despite the complexity induced by mixed
modes. Mode identification is derived by looking at the closest
ridge. In particular, we provide a straightforward determination
of the mode radial orders, which were previously unknown. Radial
eigenfrequencies are located at:
\begin{equation}\label{identin}
   \nu_{n,0} = \left[ n + \varepsilon (\dnumoy) \right] \; \dnumoy
\end{equation}
Ridges were already shown in previous works. While
\cite{2009Natur.459..398D} and \cite{2010A&A...509A..73C} looked
at single stars separately, \cite{2010ApJ...713L.176B} and
\cite{huber2010} used manual fine-tuning of the large separation
to align the radial modes of a large sample of stars. However, the
radial modes were identified in only one third of the spectra by
\cite{huber2010}, but they also showed the ridges with varying
$\varepsilon$ in the folded and collapsed power spectrum.

In most regions of the oscillation spectra  we observe the
presence of both radial and non-radial modes. Realization noise
causes the height of the individual modes to show considerable
variability, but on average, the ratio between the dipole and
radial mode height is approximately independent of $\numax$.
Although, at very low $\numax$ there is some reduction in the
strength of the dipole mode. We also make clear that the larger
spread of the ridges corresponding to dipole modes
(Fig.~\ref{identi}) is due to the presence of many mixed modes, as
already noticed \citep{2009A&A...506...57D,2010ApJ...713L.176B}.
The universal pattern makes it easier to identify them  opening up
the possibility of exploring the conditions in the inner layers of
the red giants.


Despite their low amplitudes and the resulting poor signal,
$\ell=3$ modes have been detected in Kepler data on red giants
\citep{2010ApJ...713L.176B,huber2010}. Our results represent the
first such detection in CoRoT data. Their identification gives
access to the fine structure of the oscillation spectra, as modes
of different angular degree probe different depths within the
star. Their detection and complete characterization will first be
derived from the universal pattern, then the small differences to
this pattern will be exploited to characterize in detail a given
object \citep{2010arXiv1009.1024M}.


More than 75\,\% of the red-giant candidates with brighter
magnitude than $m\ind{R}=13$ observed with CoRoT show solar-like
oscillations. In the remainder, we observe a large proportion of
classical pulsators or of giants with a so large radius that the
oscillations occur at a too low frequency for a positive
detection. In a very limited number of cases at very low
frequency, the possible confusion between radial and dipole modes
is not clearly solved. This confusion increases toward dimmer
targets with lower quality time series. Among the positive
detection of bright stars, we did not observe any outliers when
performing the correlation with the universal pattern. For this
procedure to be effective, we require that the modes have a
significant height-to-background ratio. Hence for all high
signal-to-noise targets \citep{2010A&A...517A..22M} we are able to
derive corrected values for the large frequency spacing.


It is recognized that the majority of the red giants in the CoRoT
field of view are in their post-flash helium-burning phase
\citep{2009A&A...503L..21M}. In terms of stellar evolution, the
demonstration of the universal regular pattern of red giants
proves that these red giants have similar and homologous interior
structures. On the other hand, despite the agreement of the fit in
$\varepsilon$ with the Solar value, we have verified that the
method does not work with subgiants or main-sequence stars
\citep{2009A&A...506...51B,2009A&A...507L..13B,2010ApJ...713..935B,2010arXiv1003.4368D}.
We explain this by the wide range of evolutionary phases
pertaining outside the red clump, which certainly will cause a
complex dependence of $\varepsilon$ with more parameters than just
the large separation. This disagreement reinforces the homogenous
properties of red-giant stars. The clearly observed pattern
confirms that, on average, the linewidths of the modes are
significantly smaller than the separation of the even mode pairs
and hence makes short mode lifetimes unlikely
\citep{2004SoPh..220..207S}. From the quasi-uniform width of the
ridges (Fig.~\ref{identi}), we can see that the lifetimes of the
modes increase significantly with decreasing large separation,
contrary to \cite{huber2010}. It also suggests that very complex
oscillation spectra previously observed may be an artefact of
noise.

\begin{figure}
\includegraphics[width=8.8cm]{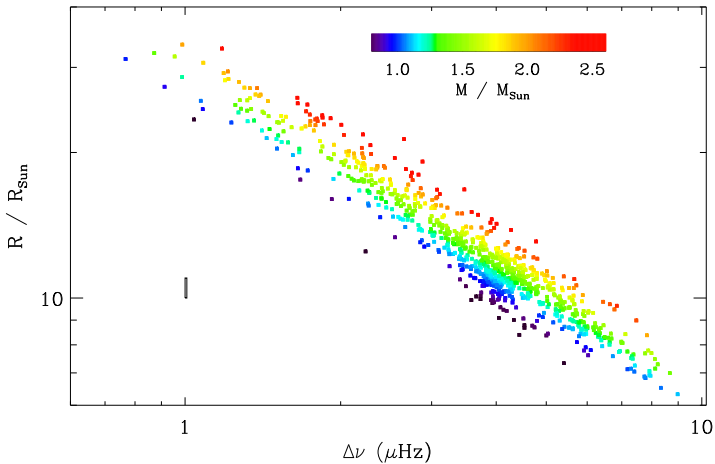}
\caption{Variation in stellar mass and radius as a function of the
large separation $\dnumoy$. The vertical line corresponds to the
error box in large separation and radius. \label{masserayon}}
\end{figure}

Finally, the better determination of the large separation
$\dnumoy$ helps us to enhance the accuracy of the estimates of the
stellar mass and radius as done in \cite{2010A&A...517A..22M},
with typical error-bars of 12 and 5\,\% respectively, instead of
20 and 8\,\% (Fig.~\ref{masserayon}). This accuracy, achieved
without the need of stellar modeling \citep{kallinger2010},
constitutes an important progress compared to the current
photometric determination and demonstrates the power of
asteroseismology.

\section{Conclusion}

We have shown with CoRoT observations that the red-giant
oscillation spectrum is very regular and can be described by its
underlying universal pattern. This was modelled in parallel by
\cite{2010ApJ...721L.182M}. As a consequence, the precise measures
of the large separation and the scaling relation of the parameter
$\varepsilon$ allow us to provide an unambiguous detection of the
radial orders and angular degrees of the modes. Since the method
is able to mitigate the realization noise, we consider it to give
the most precise determination of the large separation available.

It remains important to interpret the physical meaning of the
scaling law for the term $\varepsilon$ in the Tassoul expression.
We will have to disentangle the contributions of the surface and
of the inner region. This will require an investigation of the
term $\varepsilon$ in the context of the second-order corrections
of the Tassoul development. Very long-duration observations with
\emph{Kepler} will help for this task.

Despite the uniform aspect of the oscillation spectra, many
differences invisible in the global approach are revealed by a
detailed analysis of each individual spectrum, as, for example,
the modulation with frequency of the large separation
\citep{2010A&A...517A..22M}. Study of this variation will give
access to the most accurate analysis of the red-giant interior
structure. This summarizes the power of asteroseismology: first,
the regular pattern provides the identification of the individual
modes; second, the difference to this regular pattern unveils the
detailed interior structure. We are confident that the shifts to
the regular pattern will be explained by mass, age and/or
metallicity effects.

Similar analysis can be performed for oscillations in subgiants
and solar-like stars
\citep{2008Sci...322..558M,2008A&A...488..705A}. Due to the large
variety of evolutionary stages among those stars, we expect the
Tassoul parameter $\varepsilon$ to depend on more than the large
separation.


\begin{acknowledgements}
This work was supported by the Centre National d'\'Etudes
Spatiales (CNES). It is based on observations with CoRoT. The
research has made use of the Exo-Dat database, operated at
LAM-OAMP, Marseille, France, on behalf of the CoRoT/Exoplanet
program.
KB acknowledges financial support through a postdoctoral
fellowship from the Subside f\'ed\'eral pour la recherche 2010,
University of Li\`ege.
FJGP acknowledges FCT's grant SFRH/BPD/37491/2007. YE and SH
acknowledge financial support from the UK Science and Technology
Facilities Council (STFC). The research leading to these results
has received funding from the European Research Council under the
European Community's Seventh Framework Programme
(FP7/2007--2013)/ERC grant agreement n$^\circ$227224 (PROSPERITY),
as well as from the Research Council of K.U.Leuven grant agreement
GOA/2008/04.

\end{acknowledgements}

\bibliographystyle{aa} 
\bibliography{biblio_tu}

\end{document}